# The effect of the interface energy on pattern selection in alloy solidification: A phase-field study

Fengyi Yu[a], Qiaodan Hu[a,*], Jianguo Li[a]

[a] Shanghai Key Laboratory of Materials Laser Processing and Modification, School of Materials Science and Engineering, Shanghai Jiao Tong University, Shanghai 200240, PR China

[*] Corresponding author, E-mail address: qdhu@sjtu.edu.cn, Tel: 0086-21-54744246

## Abstract

A thorough understanding of pattern selection is necessary for the control of solidification structures, which are dissipative structures created by irreversible processes. In this paper, we simulate solidification evolution with different Preferred Crystallographic Orientations (PCOs) through the Phase-Field model. Then we study the effect of solute segregation on the interface energy, as well as the influence of the interface energy on the pattern selection. At the initial stage, the solute segregation influences the interface energy, determining the instability of the planar interface. During the detailed evolution of the Planar-Cellular-Transition (PCT), the surface stiffness dominates this stage. At the PCT stage, high degree of solute segregation refers to the low interface energy, resulting in the appearing of the sidebranches behind the tip of the primary dendrites. At the steady-state stage, the overall propagation velocities of the interfaces are the same, while the tip velocities are different in the simulations with different PCOs. The different tip velocities give rise to the different morphological evolution of the interfaces. The viewpoints of the whole dissipative system and the local free energy are discussed, respectively. This paper demonstrates the effect of solute segregation on the interface energy, as well as the influence of the interface energy on the pattern selection.

## Keywords

Interface energy; Pattern selection; Alloy solidification; Phase-field method.

The mechanical properties of as-solidified parts are determined by the solidification structures. The precise control of structures requires a thorough understanding of solidification dynamics. Since different physical processes interact with each other at different scales in solidification [1-3], the investigation of solidification dynamics has been a long standing challenge. From the mesoscale viewpoint, solidification patterns result from the interaction between the interfacial processes and the transport processes of heat and mass [3-5]. By exchanging heat and mass with the environment, solidification patterns are formed out of equilibrium, which are dissipative structures created from irreversible processes [1,2]. The dissipative structures are represented by the morphological evolution of the interface, resulting in complex solidification structures, determining the defect formation and other properties of the components.

Due to the importance of solidification structures, researchers developed various theoretical models to describe the evolution of solidification patterns. For planar interface instability, the descriptions go through the Constitutional Supercooling (CS) theory [6], Mullins-Sekerka (MS) analysis [7,8] to Warren-Langer (WL) [9] model. The theoretical predictions of incubation time and average wavelength of planar instability consist with experimental observations [10,11]. For dendritic growth in undercooled melt, starting from the Ivantsov solution [12], the theories include the maximum velocity principle [13], marginal stability hypothesis [14], microscopic solvability condition theory [15,16], and the interfacial wave theory [17]. The theoretical models could identify the important parameters determining the evolution. They can also be used as benchmark solutions to test the numerical models. However, these models involve approximations and simplifications, resulting from the constraint that the solutions can only apply under simple conditions. Hence the theoretical models can hardly handle the complex morphologies of interfaces and the interfacial effects.

As complementation to the theoretical models, the numerical models could solve the equations under complex conditions. As one of the most powerful numerical methods, the Phase-Field (PF) method combines the insights of thermodynamics and the dynamics of transport process, which has solid physical foundations [18,19]. Moreover, since it avoids the shape error caused by tracking interface in computation, the PF method has high numerical accuracy [18,19]. As for alloy solidification, with the help of thin-interface asymptotic analysis [20,21] and "Anti-Trapping Current" (ATC) term [22,23], the PF model can predict the structures quantitatively, whose results agree well with the experimental observations. Moreover, since it can capture the complex morphologies and the characteristic parameters at the interface, the PF model can represent the interaction between the interfacial processes and transport processes accurately, suitable for investigating the solidification patterns induced by the interfacial processes.

In this paper, we study the effect of the interface energy on pattern selection in alloy solidification. Through the quantitative PF model, the directional solidification with different Preferred Crystallographic Orientations (PCOs) is simulated. Based on the results, we investigate the influence of solute segregation on the interface energy, as well as the effect of the interface energy on the pattern selection.

**1. Phase field model**

We adopt the quantitative PF model for alloy solidification [22,23], with solute diffusion in the solid [24]. The following is a brief introduction.

In the PF model, a scalar variable $\phi(\mathbf{r}, t)$ is introduced to identify the phase, where $\phi = +1$ reflects the solid phase, $\phi = -1$ reflects the liquid phase, and intermediate values of $\phi$ reflects the S/L interface. Since $\phi$ varies smoothly across the interface, the sharp interface becomes diffuse and the phases turn into a continuous field, i.e., phase field $\phi(\mathbf{r}, t)$.

For the solute field, the composition $c(\mathbf{r}, t)$ is represented through the supersaturation field $U(\mathbf{r}, t)$:

$$U = \frac{1}{1-k}\left(\frac{2kc/c_\infty}{1+k-(1-k)\cdot\phi}-1\right) \tag{1}$$

where k is the solute partition coefficient, $c_\infty$ is the average solute concentration.

In alloy solidification, the ATC term can recover the local equilibrium at the interface [22], as well as eliminating the spurious effects when the interface width is larger than the capillary length. The ATC term with solute diffusion in solid is given by [24]:

$$\vec{j}_{at} = -\frac{1-k\cdot D_S/D_L}{2\sqrt{2}}\left[1+(1-k)U\right]\frac{\partial\phi}{\partial t}\frac{\vec{\nabla}\phi}{\left|\vec{\nabla}\phi\right|} \tag{2}$$

where $D_S$ and $D_L$ are the diffusion coefficients in the solid and liquid, respectively. $\partial\phi/\partial t$ reflects the rate of solidification, $\nabla\phi/|\nabla\phi|$ is the unit length along the normal direction of S/L interface.

In directional solidification, the so-called "frozen temperature approximation" is adopted,

$$T(z,t) = T_0 + G(t)\left(z - z_0 - \int V_P(t)\,dt\right) \tag{3}$$

where $T_0 = T(z_0,0)$ is a reference temperature, $G(t)$ and $V_P(t)$ are the thermal gradient and pulling speed. The frozen temperature approximation is on the basis of the following assumptions: (1) The latent heat is ignored, i.e., the temperature field is undisturbed by interfacial evolution. (2) There is no flow in the liquid, which is consistent with the assumption that the densities of the solid and liquid are equal [25].

Finally, the governing equations of the phase field and supersaturation field are given by [22,23]:

$$
\begin{aligned}
& a_s^2(\hat{n})\left[1-(1-k)\frac{z-z_0-\int V_P(t)\,dt}{l_T}\right]\frac{\partial\phi}{\partial t}= \\
& \quad \nabla\cdot\left[a_s^2(\hat{n})\vec{\nabla}\phi\right]-\partial_x\left(a_s(\hat{n})\cdot a_s'(\hat{n})\cdot\partial_y\phi\right)+\partial_y\left(a_s(\hat{n})\cdot a_s'(\hat{n})\cdot\partial_x\phi\right) \\
& \quad +\phi\left(1-\phi^2\right)-\lambda\left(1-\phi^2\right)^2\left[U+\frac{z-z_0-\int V_P(t)\,dt}{l_T}\right]
\end{aligned}
\tag{4}
$$

$$
\left(\frac{1+k}{2}-\frac{1-k}{2}\phi\right)\frac{\partial U}{\partial t}=\nabla\cdot\left[\overline{D}_L\cdot q(\phi)\cdot\vec{\nabla}U-\vec{j}_{at}\right]+\frac{1}{2}\left[1+(1-k)U\right]\frac{\partial\phi}{\partial t}
\tag{5}
$$

where,

$$l_T=\frac{\Delta T_0}{G(t)}=\frac{|m|c_\infty(1-k)}{kG(t)}$$

$$\overline{D}_L=D_L/\left(W^2/\tau_0\right)$$

$$q(\phi)=\left[kD_S+D_L+k\left(D_S-D_L\right)\phi\right]/2D_L$$

$$a_s(\hat{n})\equiv a_s\left(\theta+\theta_0\right)=1+\varepsilon_4\cos 4\left(\theta+\theta_0\right)$$

In the equations, $l_T$ is the thermal length, where m is the slope of liquidus line in the phase diagram. $a_s(n)$ is the four-fold anisotropy function in a 2D system, where $\varepsilon_4$ is the anisotropy strength, $\theta$ the angle between the normal direction of interface and the y-axis, $\theta_0$ is the intersection angle between the PCO of grain and the y-axis. $q(\phi)$ is an interpolation function determining the varied diffusion coefficient across the domain.

After ignoring the effect of kinetic undercooling, the calculation parameters in the governing equations could be linked to the physical qualities by the expressions: $W = d_0\lambda/a_1$ and $\tau_0 = a_2\lambda W^2/D_L$, where W and $\tau_0$ represent the interface width and relaxation time, which are the length scale and time scale, respectively. In the expressions, $a_1 = 5\sqrt{2}/8$ and $a_2 = 47/75$, $\lambda$ is the coupling constant, $d_0 = \Gamma/|m|(1-k)(c_\infty/k)$ is the chemical capillary length. $\Gamma = \gamma_{sl}T_f/(\rho_s L_f)$ is the Gibbs-Thomson coefficient, where $\gamma_{sl}$ is S/L interface energy, $T_f$ is the melting point of pure solvent and $L_f$ is the latent heat, respectively.

The material parameters of Al-2.0wt.%Cu, regarded as a dilute binary alloy, for the PF simulations are shown in Table 1 [26,27].

Table 1. The material parameters of Al-2.0wt.%Cu for the PF simulation [26,27]

| **Symbol** | **Value** | **Unit** |
|---|---|---|
| Liquidus temperature, $T_L$ | 927.8 | K |
| Solidus temperature, $T_S$ | 896.8 | K |
| Diffusion coefficient in liquid phase, $D_L$ | $3.0\times10^{-9}$ | $m^2/s$ |
| Diffusion coefficient in solid phase, $D_S$ | $3.0\times10^{-13}$ | $m^2/s$ |
| Equilibrium partition coefficient, k | 0.14 | / |
| Alloy composition, $c_\infty$ | 1.0, 1.5, 2.0 | wt.% |
| Liquidus slope, m | -2.6 | K/wt.% |
| Gibbs-Thomson coefficient, Γ | $2.4\times10^{-7}$ | K·m |
| Anisotropic strength of surface energy, $\varepsilon_4$ | 0.01 | / |

In the computation, the most important calculation parameter is the interface width W. The accuracy of the simulation increases with the decrease of W, while the computational cost increases dramatically with the decrease of W [20,21]. The thin interface limitation makes W just need to be one order of magnitude smaller than the characteristic length of the structure [23,28]. As for alloy solidification, the characteristic length is $L_C \sim\sqrt{d_0 * D_L/v_{tip}}$ [25], hence we set W to be 0.16μm in this paper. In the computation, the periodic boundary conditions were loaded for the phase field and supersaturation field along the Thermal Gradient Direction (TGD). The time step size was chosen below the threshold of numerical instability for diffusion equation, i.e., $\Delta t < (\Delta x)^2/(4D_L)$. We used fixed grid size $\Delta x = 0.8W$ and time step size $\Delta t = 0.012\tau_0$.

Moreover, to consider the infinitesimal perturbation of thermal noise on the S/L interface, a fluctuating current $J_U$ is introduced to the diffusion equation. By using the Euler explicit time scheme, we have:

$$U^{t+\Delta t} = U^t + \Delta t\left(\partial_t U - \vec{\nabla}\cdot\vec{J}_U\right) \tag{6}$$

The components of $J_U$ are random variables obeying a Gaussian distribution, since it has the maximum entropy relative to other probability distributions:

$$\left\langle J_U^m\left(\vec{r},\vec{t}\right) J_U^n\left(\vec{r}\,',\vec{t}\,'\right)\right\rangle = 2D_L q\left(\psi\right) F_U^0 \delta_{mn}\delta\left(\vec{r}-\vec{r}\,'\right)\delta\left(t-t'\right) \tag{7}$$

During the numerical simulation, the discretized noise in 2D becomes [29,30]:

$$\vec{\nabla}\cdot\vec{J}_U \approx \left(J^n_{x,i+1,j} - J^n_{x,i,j} + J^n_{y,i,j+1} - J^n_{y,i,j}\right)/\Delta x \tag{8}$$

In addition, the constant noise magnitude $F_u^0$ is defined as [29,30]:

$$F_U^0 = \frac{kv_0}{(1-k)^2 N_A c_\infty} \tag{9}$$

$F_U^0$ is the value of $F_U$ for a reference planar interface at temperature $T_0$, where $v_0$ is molar volume of the solute atoms, and $N_A$ is the Avogadro constant.

Finally, the program code of PF simulation was written by C++ and executed on the platform of π 2.0 cluster, supported by the Center for High Performance Computing at the Shanghai Jiao Tong University (SJTU). The explicit Finite Difference Method (FDM) was used for solving the governing equations, and the Message Passing Interface (MPI) parallelization was used for improving the computational efficiency.

**2. Results and discussion**

We adopt the dynamic solidification parameters in the simulation. The thermal gradient G is constant $10^5$K/m, while the pulling speed $V_P$ increases from 0 to a fixed value 300μm/s, for which the increase time is 2.0s. When interfacial curves exist, the PCO of crystal determines the surface energy and surface stiffness, shown in equations (10-11), respectively.

$$\gamma_{sl} = \gamma_{sl}^0 \left[1 + \varepsilon_4 \cos 4(\theta + \theta_0)\right] \tag{10}$$

$$\Psi_{sl} = \gamma_{sl} + \frac{d^2\gamma_{sl}}{d\theta^2} = \gamma_{sl}^0 \left[1 - 15\varepsilon_4 \cos 4(\theta + \theta_0)\right] \tag{11}$$

By adjusting the surface energy and/or surface stiffness, the PCO affects the solidification evolution. In this section, the simulations of crystals with different PCOs are carried out, for which the PCOs are set to be 0°, 5°, 10°, 15°, 20°, 25°, 30°, 35°, 40°, and 45°, respectively. The computational domain is 2000×2000 grids, corresponding to 256.0μm×256.0μm in the real unit. It takes 20 hours using 40 cores to finish one job.

The evolution of the characteristic parameters is shown in Figure 1, including the solute concentration ahead of the interface $c_0$ and the instantaneous velocity of the interface $V_{tip}$. The $c_0$ increases with time, so does the $V_{tip}$. As time goes on, the planar instability occurs, represented by the transformation from the planar to the cellular, shown in Figure 2. Due to the cellular shape, the $V_{tip}$ increases sharply, shown by the sharp increment of the $V_{tip}$ curves in Figure 1(b1)-(b2). Meanwhile, the solute at the S/L interface should satisfy the conservation law. At the crossover time of the instability, the solute still accumulates ahead of the interface, shown by the limited increase of the $c_0$ curves after the crossover time in Figure 1(a1)-(a2). After the cellular appearing, rather than diffusing along the pulling direction of the planar interface, the solute could diffuse along multiple directions from the cellular tip to the liquid. As a result, the solute concentration starts to decrease, shown by the decrease of $c_0$ curves in Figure 1(a1)-(a2). In conclusion, the sharp increment of the instantaneous velocity reflects the onset time of the planar instability, and the peak of the solute concentration reflects the completion of the cellular appearing.

Due to the importance of the planar instability, the criterion of whom needs to be determined. Combining the revised CS theory, in equation (12) [8], and the time-dependent concentration gradient ahead of the interface, in equation (13) [9], the criterion of interface stability based on $V_{tip}$ is given by equation (14).

$$m_l G_{Cl}^* \leq G \tag{12}$$

$$G_{Cl}^* = \left.\frac{\partial c_0}{\partial z}\right|_{z0} = \frac{V_{tip}\left(1-k\right)c_0\left(z_0,t\right)}{-D_L} \tag{13}$$

$$V_{tip} \cdot c_0\left(z_0,t\right) \leq 134.2\times10^{-6} \tag{14}$$

In Figure 1(a2), the solute concentration at the crossover time is 5.7%. According to equation (14), the critical velocity of interface instability is 23.5μm/s, which is not consistent with the PF simulation in Figure 1(b2), where $V_{tip}$ is 116.7μm/s. That is, the tip velocity cannot be the criterion for the interface stability.

According to the literature [31], in alloy solidification, the cooling rate ($G^*V_P$) dominates the overall propagation speed of the interface, to maintain the local thermodynamic equilibrium. The solute segregation determines the stability of interface, by changing the excess free energy and corresponding interface energy. Since the same solidification parameters are used in the simulations, the $c_0$ curves overlap with each other completely before the crossover time of the planar instability, in Figure 1(a1), so do the $V_{tip}$ curves in Figure 1(b1). The results consistent with the literature [30], i.e., the surface energy and its anisotropy do not affect the solute diffusion and planar growth. Under the same solidification conditions, at the planar growth stage, the evolution of solute segregation is the same in the simulations with different PCOs, resulting in the same crossover times of the planar instability.

On the other hand, although the crossover times of the instability are the same in the simulations with different PCOs, the detailed characteristics differ with each other, shown in Figure 1(a2)-(b2). The differences include the extreme values of $c_0$ and $V_{tip}$, as well as the amount of time of the Planar-Cellular-Transition (PCT). The distinctions result from the different PCOs of grain, as well as the surface energy and/or surface stiffness. There are two common rules for the selection of growth direction: the maximum surface energy and minimum surface stiffness. For the cubic crystal, the rule of maximum surface energy means the crystal will seek to minimize the total surface energy by creating higher curvature in the <100> direction, while the rule of minimum surface stiffness means the crystal prefer to grow in the direction where the surface presents the smallest resistance to being deformed [25]. For the PCT, it is hardly for the crystal to directly create large curvature in the <100> direction from the planar with zero-curvature, due to the great energy barrier caused by the curvature difference. That is, the maximum surface energy rule is not suitable for the PCT. We move on to the minimum surface stiffness rule, in equation (11), the surface stiffness is expressed by $\Psi_{sl} = \gamma^0_{sl}[1-15\varepsilon_4\cos4(\theta+\theta_0)]$, where $\theta$ is the angle between the normal direction of the interface and the y-axis. At the planar growth stage, the value of $\theta$ is infinitesimal, hence the surface stiffness could be simplified as $\Psi_{sl} = \gamma^0_{sl}[1-15\varepsilon_4\cos4\theta_0]$, indicating the value of surface stiffness increases from $\theta_0 = 0°$ to $\theta_0 = 45°$. According to the minimum surface stiffness rule, the crystal prefer to grow in the direction where the surface presents the smallest resistance to being deformed. As a result, the cellular with $\theta_0 = 0°$ is the easiest to appear, while the cellular with $\theta_0 = 45°$ is the hardest to appear, shown in Figure 1(a2), where the increases rate of tip velocity decreases from $\theta_0 = 0°$ to $\theta_0 = 45°$ after the crossover time.

After the planar instability, the cellular prefer to growth along their <100> direction, for which it takes time from the perturbation of the interface to the cellular shape. From t = 1.558s (in Figure 2) to t = 1.993s (in Figure 3), solute concentration increases to a maximum value and then decreases, shown in Figure 1(a2). The different tip velocities, caused by the different interface stiffness, give rise to different degrees of solute segregation ahead of the interface. On the other hand, due to the multiple directions of solute diffusion after the cellular appearing, the solute segregation start to decreases. The lower tip velocity means the longer time for the PCT, corresponding to the longer time of the cellular appearing and the higher degree of solute segregation, in Figure 1(a2)-(b2), from $\theta_0 = 0°$ to $\theta_0 = 45°$. In turn, the higher degree of solute segregation corresponds to the lower interface energy, as well as the lower interface stiffness. Since the rule of minimum surface stiffness is responsible for the PCT, the higher degree of solute segregation means the lower interface stiffness. The peak values of tip velocity decreases from $\theta_0 = 45°$ to $\theta_0 = 0°$, shown in Figure 1(a2)-(b2). It should be noted, from t = 1.558s to t = 1.993s, a few sidebranches appear behind the tip of primary dendrite, in Figure 3, due to the decrease of interface energy induced by solute segregation. Subsequently, with the solute concentration decreasing, fewer sidebranches grow out.

It should be pointed out, after the cellular appearing, the stochastic factors (noise) make the cellular have different shapes and growth speeds, resulting in temporary competition between the cellular, represented by the increasing arm spacing and the decreasing number of the cellular from the bottom to the top in Figure 3. Meanwhile, the distortions of interfacial morphology are observed in Figure 3, which can be regarded the effect of internal stress. From both the practical (measurements) and theoretical (thermodynamics) point of view, the surface energy $\gamma_{sl}$ and any surface stress related to it are defined at equilibrium. Because $\gamma_{sl}$ is the work of creating a new surface at equilibrium, including surface distortions of all kinds. Thus, internal stresses are already subsumed into the surface energy $\gamma_{sl}$, and no additional effect [32].

As time goes further, solidification turns into the steady-state growth, shown by the relatively stable curves in Figure 1(a1)-(b1) after t = 2.5s and the detailed evolution of interfacial morphology and solute field in Figure 4. At this stage, the overall propagation velocities of interface along the pulling direction are the same in the simulations. Because the dissipative structures achieve a quasi-steady state after a period of self-organization, the overall propagation velocities equal to the pulling speed. Although they have similar overall velocities, the grains with different PCOs grow with different tip velocities, in Figure 1(b1) after t = 2.5s, where the tip velocity is defined by the expression $v_{tip} = [z_0(t_2)-z_0(t_1)/(t_2-t_1)]/\cos\varphi_0$. In the expression, $\varphi_0$ is the angle between the growth direction of the primary dendrite and y-axis, here we regard $\varphi_0$ equals to $\theta_0$. According to the expression of $V_{tip}$, the grains with the larger PCOs grow with the larger tip velocities. By comparing the simulations in Figure 4, the grains with larger PCOs are more likely to grow out sidebranches, due to their different tip velocities. From the viewpoint of the whole domain, to keep the quasi-steady state of the dissipative structures, the system needs to exchange heat and mass with the environment. The larger tip velocity refers to the higher degree of non-equilibrium of the system, requiring more heat and/or mass exchange, resulting in more non-equilibrium structures (sidebranches). Hence the grains with larger PCOs are more likely to grow out sidebranches. From the viewpoint of the local domain, the onset of sidebranches can be regarded as one kind of interface instability, determined by the interface energy and its anisotropy. As shown in Figure 4, the sidebranches always appear behind the critical solute concentration, shown by the 4.5wt%Cu curves. Since the solute segregation decreases the interface energy, when the interface energy reduces to the critical level, the interface instability occurs, resulting in the onset of sidebranches.

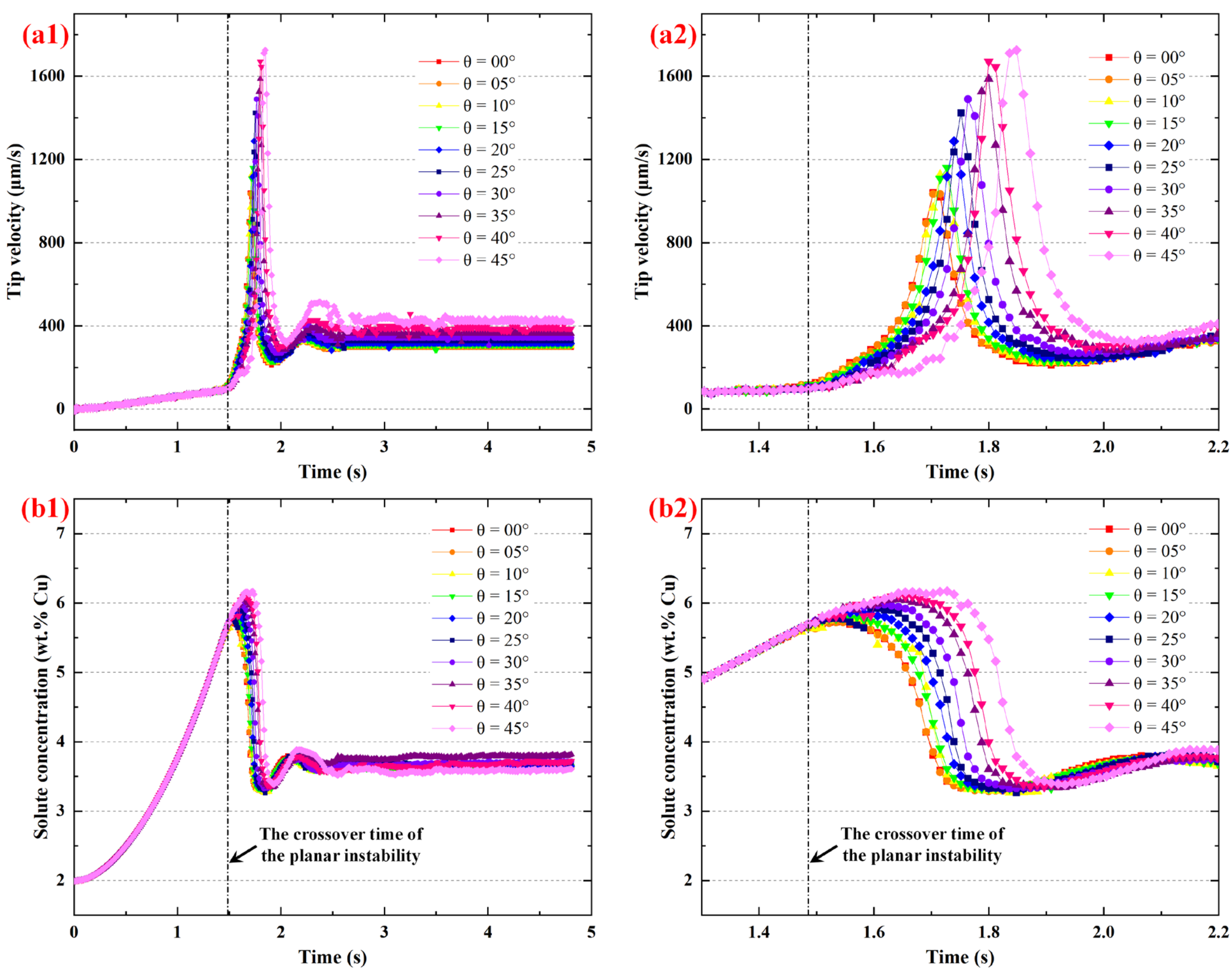

Figure 1. The evolution of characteristic parameters: (a) the solute concentration ahead of the interface and the enlarged version of (a1); (b) the instantaneous velocity of the interface and the enlarged version of (b1).

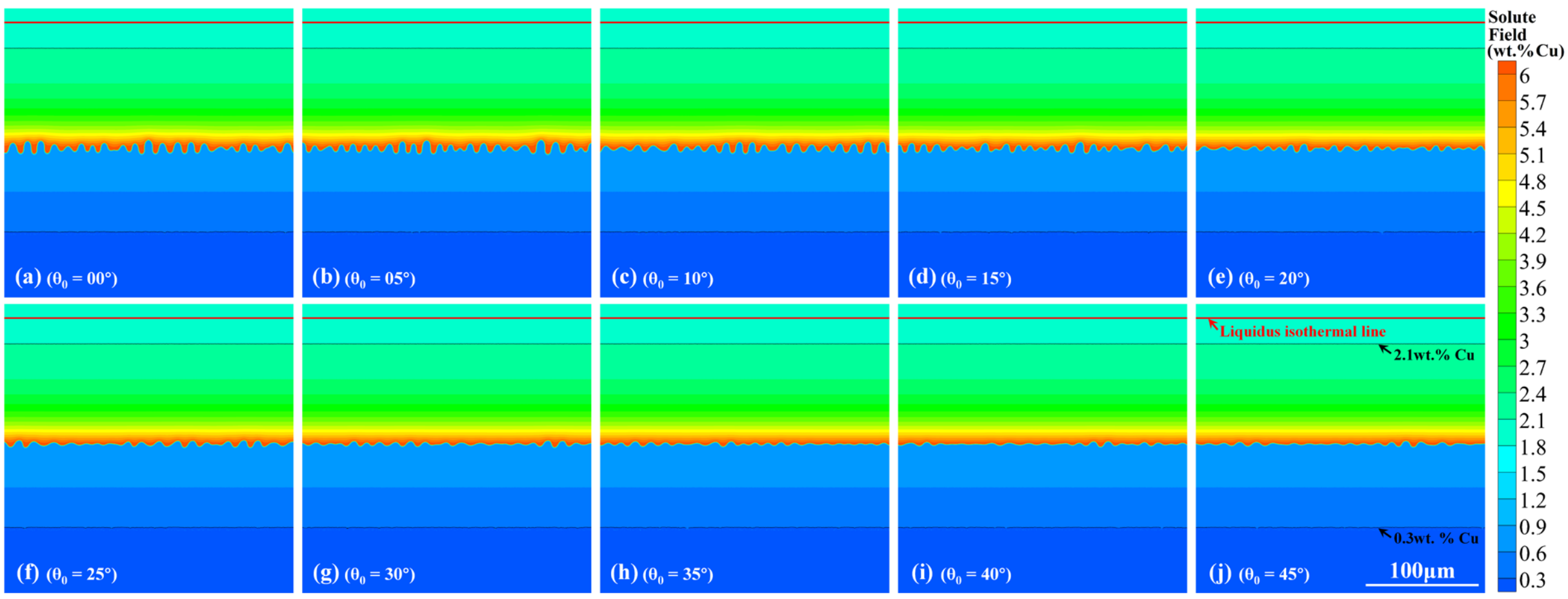

Figure 2. The evolution of the interfacial morphology and solute field with different PCOs of grain from the PF simulations (t = 1.558s)

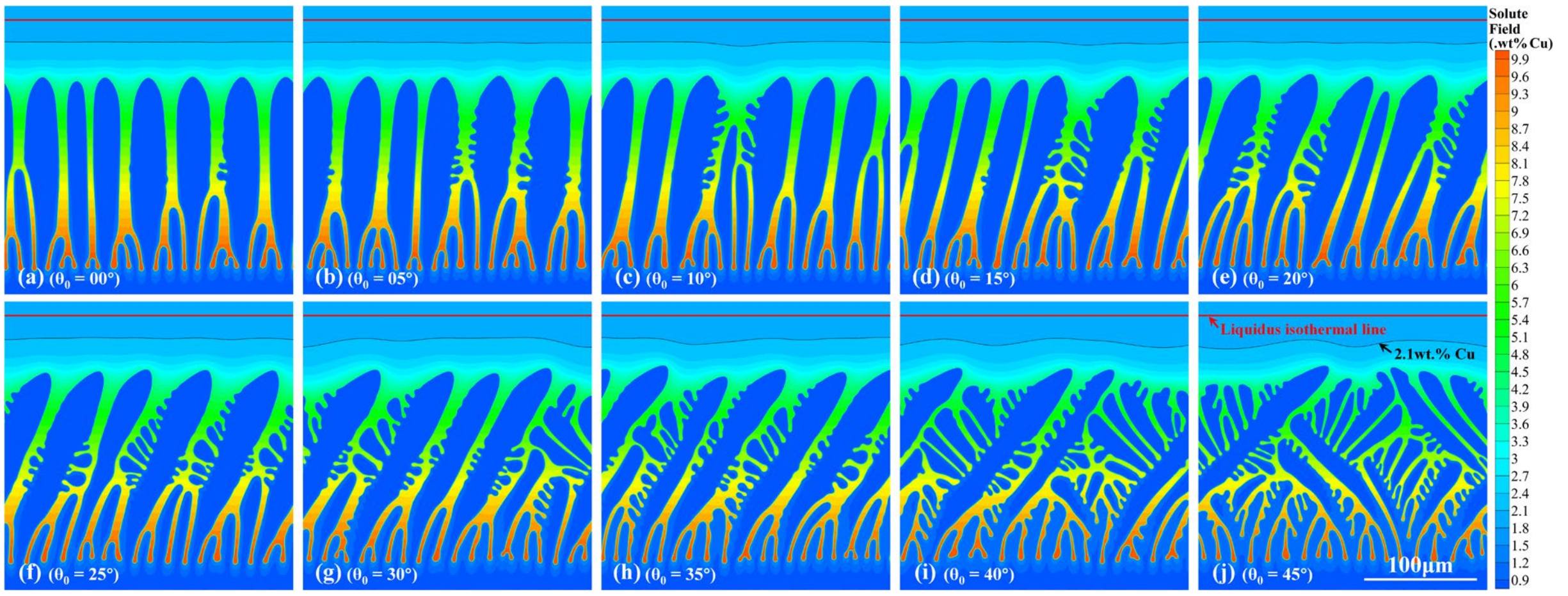

Figure 3. The evolution of the interfacial morphology and solute field with different PCOs of grain from the PF simulations (t = 1.993s)

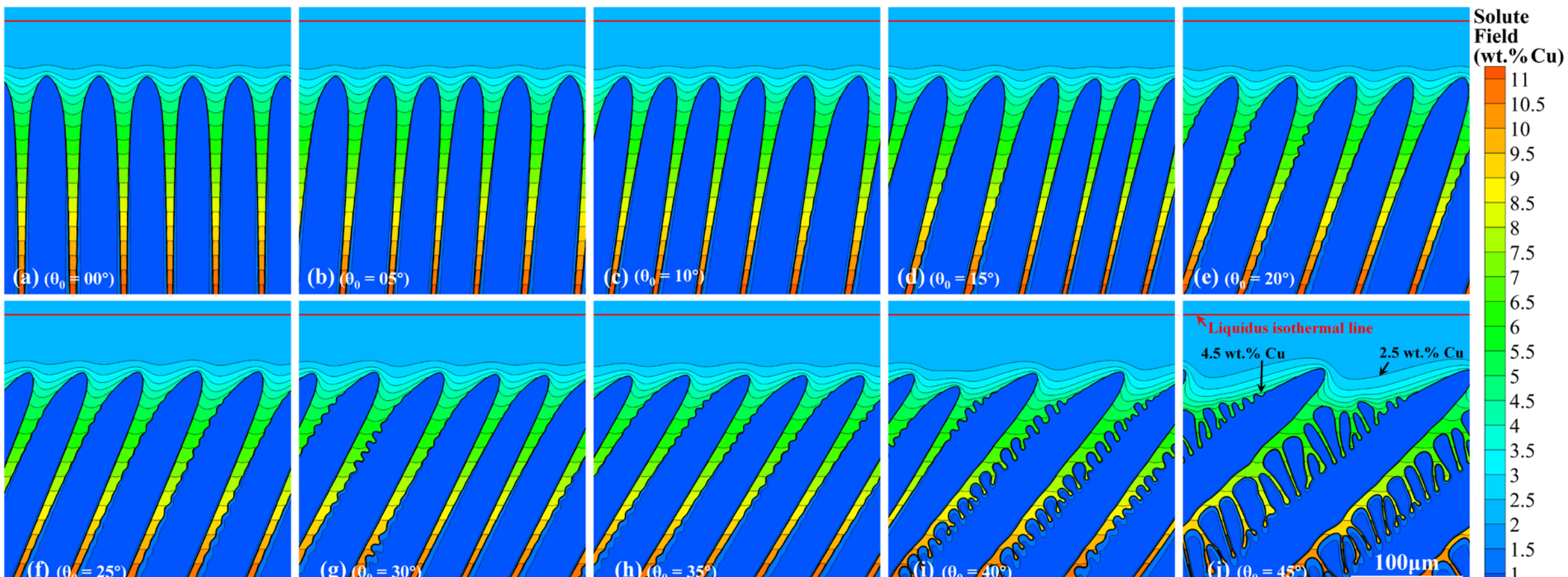

Figure 4. The evolution of the interfacial morphology and solute field with different PCOs of grain from the PF simulations (t = 4.819s)

## 3. Conclusion

In this paper, we investigate the effect of the interface energy on pattern selection in alloy solidification. By the quantitative PF model, the directional solidification with different PCOs is simulated. Based on the results, we study the effect of solute segregation on the interface energy, as well as the influence of the interface energy on the pattern selection. The following conclusions could be drawn from the study:

(1) At the initial stage, the sharp increment of the instantaneous velocity represents the onset time of the planar instability, and the peak of the solute concentration reflects the completion of the cellular appearing.

(2) The solute segregation affects the interface energy, determining the stability of the interface. At the planar growth stage, the evolution of solute segregation is the same in the simulations with different PCOs, resulting in the same crossover times of the planar instability.

(3) During the detailed evolution of the PCT, the surface stiffness dominates this transition. Specifically, the surface stiffness increases from $\theta_0 = 0°$ to $\theta_0 = 45°$, resulting in the lower tip velocities, the longer time of the PCT, the higher degree of solute segregation and the increasing peak values of the tip velocities.

(4) On the other hand, during the evolution of the PCT, high degree of solute segregation refers to the low interface energy, resulting in the appearing of the sidebranches behind the tip of the primary dendrites. Subsequently, with the solute concentration decreasing, fewer sidebranches grow out.

(5) At the steady-state growth stage, the overall propagation velocities of the interfaces are the same, while the tip velocities are different in the simulations with different PCOs. The different tip velocities result in different morphological evolution of the interfaces. From the viewpoint of the whole dissipative system, larger tip velocity reflects higher degree of non-equilibrium of the system, which requires more heat and/or mass exchange, resulting in more non-equilibrium structures (sidebranches). From the viewpoint of the local domain, sidebranches appear after the critical solute concentration, due to the fact that the solute segregation determines the interface energy.

This paper indicates the influence of the interface energy on the pattern selection, where the interface energy is determined by solute segregation. The non-equilibrium degree of the system is considered when investigating the pattern selection. As one kind of dissipative structures, irreversibility and non-equilibrium are important factors during the evolution of solidification patterns, which will be studied in the future.

**Acknowledgments**

This work is supported by the fellowship of China Postdoctoral Science Foundation (2021M692040), National Key Research and Development Program (2017YFB0305301), and National Natural Science Foundation of China-Excellent Young Scholars (51922068). The authors acknowledge the technical support from the Center for High Performance Computing at the SJTU.

**Appendix**

The derivation of equation (14) is following. The revised CS criterion [8]:

$$m_l G_{Cl}^* \leq G$$

The time-dependent concentration gradient ahead of the interface [9]:

$$G_{Cl}^* = \left.\frac{\partial c_0}{\partial z}\right|_{z0} = \frac{V_{tip}\left(1-k\right)c_0\left(z_0,t\right)}{-D_L}$$

Then, we have:

$$m_l \frac{V_{tip}\left(1-k\right)c_0\left(z_0,t\right)}{-D_L} \leq G \Rightarrow \frac{2.6\times 0.86\cdot V_{tip}\cdot c_0\left(z_0,t\right)}{3\times 10^{-9}} \leq 10^5 \Rightarrow V_{tip}\cdot c_0\left(z_0,t\right) \leq 134.2\times 10^{-6}$$

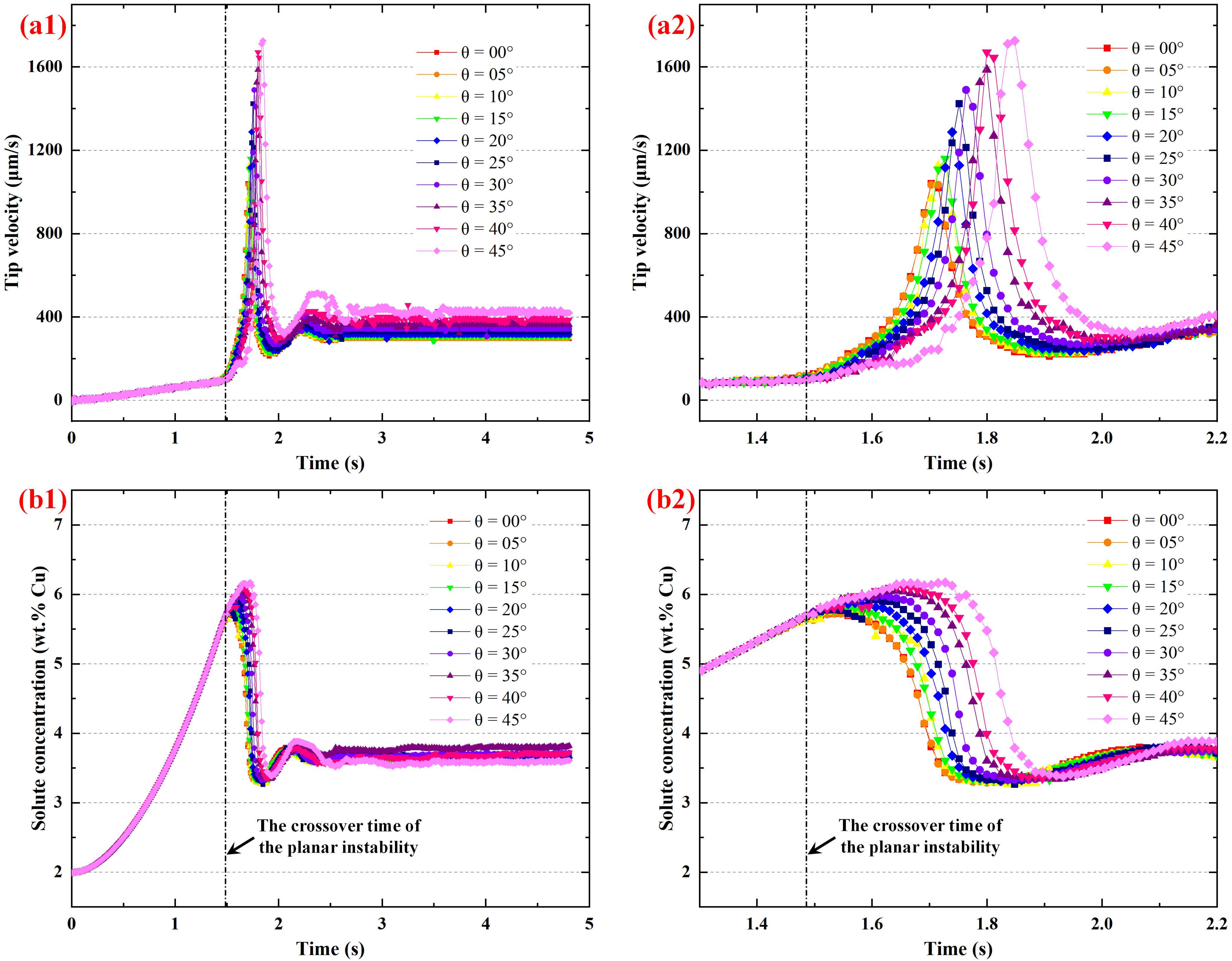

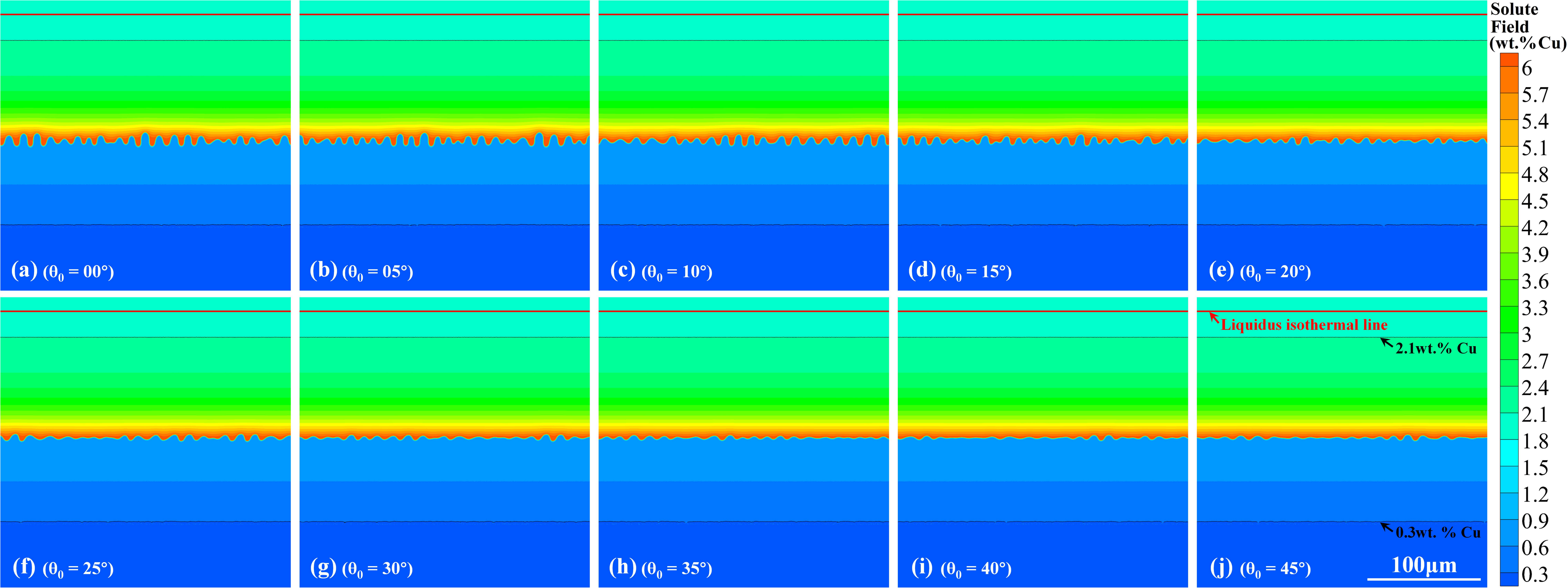

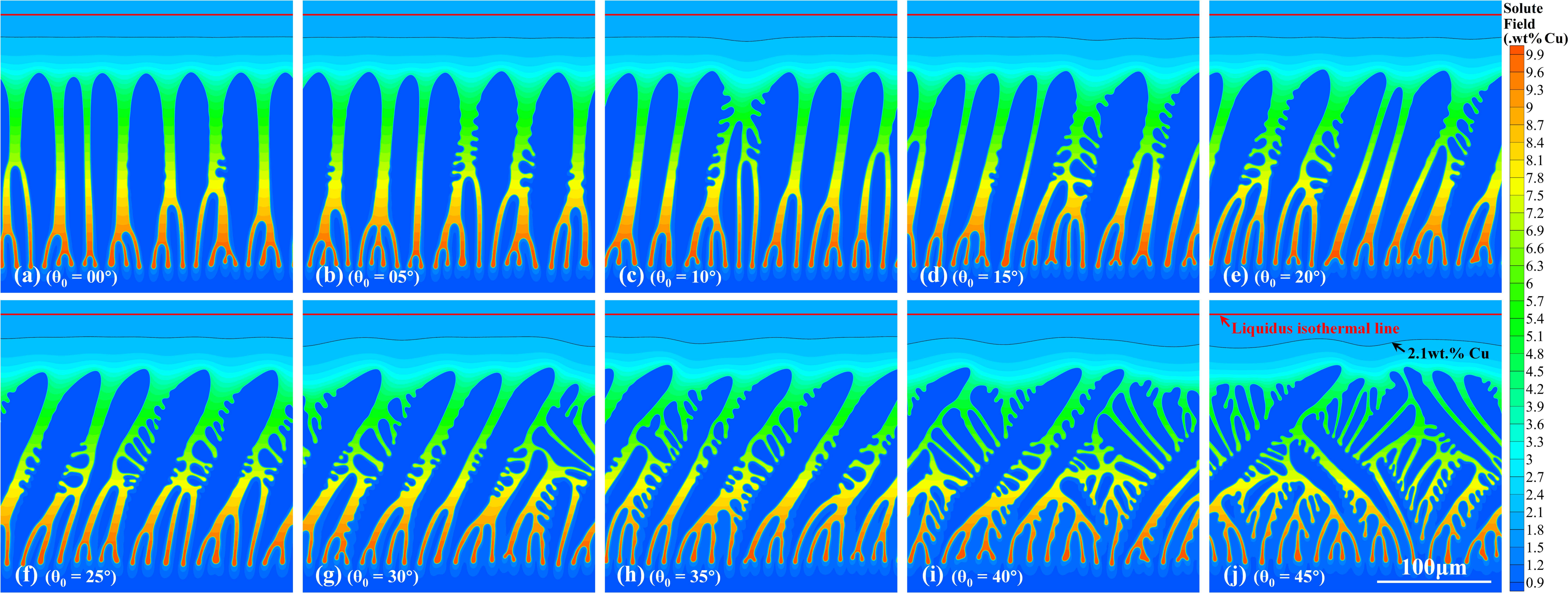

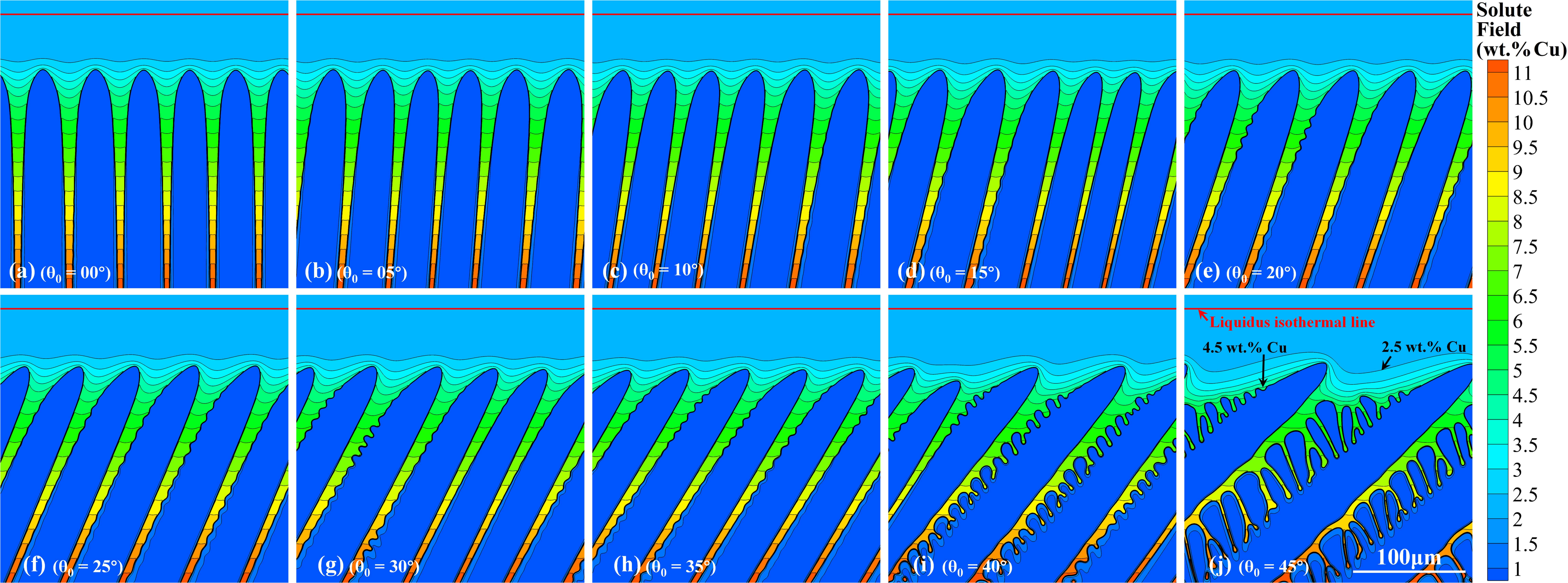